\documentclass[aps, pra, twocolumn, superscriptaddress, amsmath,  tightenlines, longbibliography]{revtex4-2}

\usepackage{dcolumn}
\usepackage{graphicx}
\usepackage{epstopdf}
\usepackage{mathrsfs}
\usepackage{subfigure}
\usepackage{booktabs}
\usepackage{amsmath,amsfonts}
\usepackage{physics}
\usepackage{dsfont}
\usepackage{amstext}
\usepackage{amssymb}
\usepackage{amsbsy}
\usepackage{bbm}
\usepackage{amsthm}
\usepackage{graphicx}
\usepackage{xcolor}
\usepackage{CJK}
\usepackage[colorlinks,urlcolor=blue,linkcolor=blue,citecolor=blue]{hyperref}
\usepackage{soul}

\usepackage{url}
\usepackage[colorlinks]{hyperref}
\hypersetup{%
	plainpages=true,
	breaklinks=true,       
	hypertexnames=false,  
	pageanchor=true,
	colorlinks=true,
	linkcolor={blue},
	citecolor={blue},
	urlcolor={blue},
	anchorcolor={black}
}

 \makeatletter

\newcommand{\Rmnum}[1]{\expandafter\@slowromancap\romannumeral #1@}
\makeatother

\begin{document}

\title{
Improve Quantum-Battery Lifetime by Electromagnetically-Induced Transparency Effect and Bound State}
    \author{Jin-Tian Zhang}
    \affiliation{%
		School of Physics and Astronomy, Applied Optics Beijing Area Major Laboratory, Beijing
 Normal University, Beijing 100875, China
	}%
 \affiliation{%
		Key Laboratory of Multiscale Spin Physics, Ministry of Education, Beijing Normal University, Beijing 100875, China
	}%

	\author{Cheng-Ge Liu}
    \affiliation{%
		School of Physics and Astronomy, Applied Optics Beijing Area Major Laboratory, Beijing
 Normal University, Beijing 100875, China
	}%
 \affiliation{%
		Key Laboratory of Multiscale Spin Physics, Ministry of Education, Beijing Normal University, Beijing 100875, China
	}%

	\author{Qing Ai}
	\email{aiqing@bnu.edu.cn}
	\affiliation{%
		School of Physics and Astronomy, Applied Optics Beijing Area Major Laboratory, Beijing
 Normal University, Beijing 100875, China
	}%
 \affiliation{%
		Key Laboratory of Multiscale Spin Physics, Ministry of Education, Beijing Normal University, Beijing 100875, China
	}%




\date{\today}

\begin{abstract}

Quantum battery (QB) is an application of quantum thermodynamics which uses quantum effects to store and transfer energy, overcoming the limitations of classical batteries and potentially improving the QB's performance.  However, due to the interaction with the external environment, it leads to decoherence and thus reduces both the lifetime and the charging efficiency of QB. Here, we propose suppressing the environmental dissipation in the energy-storage process of the QB by exploiting both the electromagnetically-induced transparency (EIT) and bound states. By constructing a hybrid system composed of a four-level atom and a coupled-cavity array, two bound states are formed in the system when the energy of the QB is in the energy band of the cavity array. Environmental dissipation is significantly inhibited due to the bound state and EIT effects, which extends the lifetime of the QB. We show that the charging efficiency of the QB is optimal when the energy of the QB is in resonance with the cavity. In addition, there is an optimal coupling strength between the two adjacent cavities, which helps to improve the performance of the QB. Our research has achieved the realization of long-term energy storage while improving the charging efficiency of QB, which has practical implications for the realization of QB.

\end{abstract}

\maketitle 

\section{Introduction} 

In recent decades, due to the application of nanomechanical electronic devices, quantum technology has gradually attracted broad interest \cite{Kishor2022PRM,Xu2020PRM}. Quantum thermodynamics focuses on the transfer and storage of energy at the nanoscale \cite{Massimiliano2009PRM,Vinjanampathy2016PC,Benenti2017RP,Matteo2022PRX,Michele2017PRE}. It studies work and entropy in a quantum context \cite{Goold2016JPA,Anders2017NJP,Ronnie2013Entropy,Mitchison2019CP}. The laws of classical thermodynamics have been reconsidered to describe heat engines and energy-storage systems based on quantum effects, breaking the limits of classical physics \cite{James2019PRL,Ronagel2014PRL,Noah2010PRL,Marlan2003Science}. As a nanoscale energy storage and conversion device, the aim of the design of the quantum battery (QB) is fundamentally different from that of conventional batteries by exploiting the properties of quantum thermodynamics \cite{Robert2013PRE,Robert2013PRE,Hovhannisyan2013PRL,Binder2015NJP}.

The QBs utilize quantum effects such as entanglement \cite{Llobet2015PRX,Marcello2019PRL,Robert2013PRE,Jonathan2002PRL,Gyhm2024AVS} and coherence \cite{Raam2015PRX,Cakmak2020PRE,Francica2020PRL,Monsel2020PRL} to achieve higher charging power and battery capacity than their classical counterparts \cite{Xue2023PRL,Work2023PRL,Shi2022PRL,Hovhannisyan2013PRL}. The basic mechanism underlying the QBs is that for example, a two-level system acts as a QB and another two-level system \cite{Salvia2023PRR,Chen2022PRE,Seah2021PRL,Andolina2018PRB} or external field acts as a charger \cite{Santos2023PRA,Xiang2023PRA,Arjmandi2022PRA,Santos2019PRE}. Here, due to their interaction, the energy can be coherently transferred between the charger and the QBs, and thus results in the charged QBs. It has been experimentally demonstrated that charging of the QBs can be achieved in various physical systems including nuclear magnetic resonance \cite{Joshi2022PRA}, superconducting circuits \cite{Zheng2022NJP,Hu2022QST}, photonic systems \cite{Maillette2023PRL} and organic microcavities \cite{Zhu2023PRL}. An efficient QB scheme is essential for its implementation and future applications, e.g. utilizing the collective-excitation mode \cite{Dario2018PRL}, and a one-dimensional dimerized XY chain as a spin QB \cite{Grazi2024PRL}. Since the realization of the QB is generally considered in the context of open quantum systems \cite{Farina2019PRB,Liu2019JPCC}, a serious problem for QB may occur. Due to the decoherence caused by the external environment, the charging efficiency of the QB will be reduced, and the stored energy will be depleted \cite{Pirmoradian2019PRA}. Various methods have been proposed to beat the dissipation, e.g. feedback control \cite{Yao2022PRE} and Floquet engineering \cite{Bai2020PRA}. However, how to improve the charging efficiency of QB while supress the decoherence caused by the external environment is an intriguing challenge.

As known to all, electromagnetically-induced transparency (EIT) is a quantum-coherence phenomenon \cite{Harris1990PRL}, which is widely observed in atomic or molecular systems. When a probe field passes a medium with resonant two-level atoms, i.e., the ground state and a lossy excited state, the transmittance of the probe field will significantly decrease due to the  absorption. However, if a control field is introduced to induce the transition between the lossy excited state and a metastable state, the probe field will pass the medium as if it is transparent. And thus it is called EIT. There are typically three kinds of configurations for the EIT, i.e., the ladder type, V-type and $\Lambda$ type \cite{Boon1999PRA}. In the EIT, there exists a dark state, i.e., the superposition of the ground state and the metastable state \cite{Wang2018PRA}. Due to the dark state being void of the lossy excited state, the probe field can pass the medium without absorption. The dark state has been applied in different physical processes, e.g. coherent energy transfer of artificial light harvesting \cite{Dong2012LSA} and perfect state transfer in optomechanical systems \cite{Wang2012PRL,Khanaliloo2015PRX}. Recently, remote charging and degradation suppression of the QB has been proposed when bound states form in the total system including the charger, the QB, and the electromagnetic field \cite{Bai2019PRL,Song2024PRL}. In this work, inspired by these discoveries, we theoretically propose a QB scheme utilizing a hybrid system composed of a four-level atom as the QB and a coupled-cavity array as the charger, which can be implemented with superconducting transmission line resonators \cite{You2003PRB,chiorescu2004Nature,wallraff2004nature,Weng2023FR} and a superconducting charge qubit \cite{sillanpaa2007Nature,you2005PT}. By applying two coherent light beams to the four-level atom, we can realize a quasi-two-level system with a dark state and the ground state for the QB. We show that the interaction between the dark state of the QB and the coupled-cavity array gives rise to two bound states in the hybrid system. Remarkably, the bound states assisted by the dark state significantly suppresses the environmental dissipation, thereby enhancing the QB's lifetime. Furthermore, our results reveal that the extractable work, i.e., the ergotropy, reaches its maximum when the dark state is resonant with the bare frequency of the cavity. We further demonstrate that the coupling strength between adjacent cavities can be optimized to enhance the performance of the QB. Unlike traditional dark-state QB \cite{Liu2019JPCC,Liu2024Molecules,Quach2020PRA,Santos2019PRE}, we propose a charging and energy-storage mechanism that integrates the effects of the EIT with the formation of bound states. The introduction of EIT does not only enhance the coherence between the QB and its environment, but also facilitates the formation of the dark. These special states including the bound states can significantly extend the lifetime of the QB while simultaneously improving its charging efficiency.

This article is structured as follows. In the next section, we will introduce our QB protocol. By using the dark state, we obtain the effective Hamiltonian of the QB interacting with the coupled-cavity array, as detailed derived in Appendix~\ref{AppendixA}. It is demonstrated that the eigen energies of the bound states determine the quantum dynamics of the dark state, with its detailed derivation shown in Appendix~\ref{AppendixB}. Then, in Sec.~\ref{sec:Discussion}, we perform numerical simulations. The relationship between the number of bound states and the energy of the dark state is explored. We use both analytical and numerical methods to verify that in the presence of two bound states, environmental dissipation can be significantly suppressed due to the combination of the bound states and the dark state. The charging efficiency of the QB is optimal when the energy of the QB is in resonance with the cavity. In Sec.~\ref{sec:Conclusion}, we summarize our main findings.

\section{Scheme}\label{Sec:Scheme}

\begin{figure}[htbp]
\centering
\includegraphics[bb=0 0 855 200,width=8.3cm]
{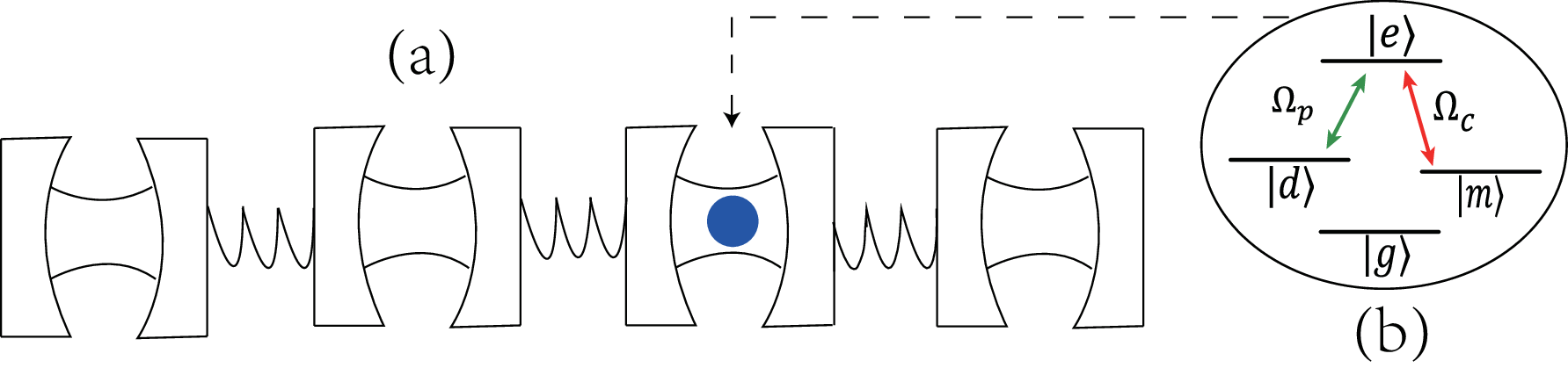}
\caption{Schematic illustration of charging QB model. (a) Structure of coupled cavities, (b)a four-level atom, where $|d\rangle$ is a metastable energy level, $|m\rangle$ is a auxiliary energy level and $|e\rangle$ is a excited state energy level. Two pulses are applied at the same time to induce the transitions of $|e\rangle \leftrightarrow |d\rangle$ and $|e\rangle \leftrightarrow |m\rangle$ with respectively Rabi frequency $\Omega_p$ and $\Omega_c$.}\label{fig:1}
\end{figure}

To study the effects of the environmental dissipation on the energy storage process of a QB, we consider the interaction between a four-level atom and a coupled-cavity array. As shown in Fig.~\ref{fig:1}, we make use of the four-level atom as a tool for energy storage, i.e., a QB, and the coupled $N$ cavities as a medium for energy transfer. The Hamiltonian of the system reads
\begin{eqnarray}
H&=&H_c+H_B+H_I,\label{eq:1}\\
H_{c}&=&\sum_{j}\omega_{0}a_{j}^{\dagger}a_{j}-\xi\sum_{j}(a_{j}^{\dagger}a_{j+1}+a_{j+1}^{\dagger}a_{j}),\label{eq:2}\\
H_{B}&=&\Omega_{d}|d\rangle\langle d|+\Omega_{e}|e\rangle\langle e|+\Omega_{m}|m\rangle\langle m|\nonumber\\
& &+\Omega_{p}e^{-i\omega_{p}t}|e\rangle\langle d|+\Omega_{c}e^{-i\omega_{c}t}|e\rangle\langle m|+\rm h.c.,\label{eq:3}\\
H_{I}&=&g_{1}a_{0}^{\dagger}|g\rangle\langle d|+g_{2}a_{0}^{\dagger}|g\rangle\langle m|+\rm h.c.,\label{eq:4}
\end{eqnarray}
where $a_{j}^{\dagger}(a_{j})$ represents create (annihilate) a photon with frequency $\omega_0$ in the $j$th cavity, $\xi$ is the coupling constant between two neighbouring resonators, $\Omega_{d}$, $\Omega_{e}$, $\Omega_{m}$ are respectively the transition frequencies of the four-level atom from $|g\rangle$ to $|d\rangle$, $|e\rangle$, $|m\rangle$. Between the energy levels $|d\rangle(|m\rangle)$ and $|e\rangle$, we apply a probe (control) light with the driving frequency $\omega_{p(c)}$ and Rabi frequency $\Omega_{p(c)}$. $g_1(g_2)$ is the coupling constant between the transition $|g\rangle\leftrightarrow|d\rangle(|g\rangle\leftrightarrow|m\rangle)$ and the $0$th cavity. In this article, we have assumed $\hbar = 1$ for simplicity. 
Here, we apply the classical drivings which are respectively in close resonance with the transitions $|e\rangle\leftrightarrows|d\rangle$ and $|e\rangle\leftrightarrows|m\rangle$. Meanwhile, the classical drivings are largely detuned with the transitions $|g\rangle\leftrightarrows|d\rangle$ and $|g\rangle\leftrightarrows|m\rangle$. As a result, in Eq.~(\ref{eq:3}) we do not consider interaction terms between $|g\rangle$ and $|d\rangle$/$|m\rangle$.
In Eq.~(\ref{eq:4}), there is no interaction between the
intermediate states and the excited state. These could be achieved by tuning the quantized field in the cavity to be resonant with the transitions $|g\rangle\leftrightarrows|d\rangle$ and $|g\rangle\leftrightarrows|m\rangle$, while largely detuned with the transitions $|e\rangle\leftrightarrows|d\rangle$ and $|e\rangle\leftrightarrows|m\rangle$.

Assuming the periodical boundary condition, by Fourier transformation, i.e., $a_{j}^{\dagger}=\sum_{k}a_{k}^{\dagger}\exp(ikj)/\sqrt{N}$, and with respect to the rotating frame $U_R=\exp(-i\omega_{e}t|e\rangle\langle e|)$, the total Hamiltonian~(\ref{eq:1}) can be rewritten as
\begin{eqnarray}
H^\prime&=&\sum_{k}\omega_{k}a_{k}^{\dagger}a_{k}+\frac{1}{\sqrt{N}}\left[\sum_{k}g_{1}(a_{k}^{\dagger}|g\rangle\langle \ d|+a_{k}|d\rangle\langle g|)\right.\nonumber\\
 & &\left.+\sum_{k}g_{2}(a_{k}^{\dagger}|g\rangle\langle m|+a_{k}|m\rangle\langle g|)\right]+H_{B}',\label{eq:5}\\
H_B'&=&\Delta_{e}|e\rangle\langle e|+\Omega_{d}|d\rangle\langle d|+\Omega_{m}|m\rangle\langle m|+\Omega_{p}|e\rangle\langle d|\nonumber\\
& &+\Omega_{c}|e\rangle\langle m|+\rm h.c.,\label{eq:6}
\end{eqnarray}
where $\omega_k=\omega_0-2\xi\cos(k)$, $\Delta_{e}=\Omega_{e}-\omega_{e}$. Because the energy levels $d$ and $m$ are metastable, we only consider that there is dissipation on the energy level $d$. By introducing a imaginary part on the energy of $|d\rangle$, i.e., $\Omega_{d}=\Omega_d'-i\kappa/2$, with $\kappa$ being the dissipation rate, we can investigate the effects of the dissipation. 
This approach is theoretically sound as long as the characteristic frequency of the Hamiltonian is larger than the dissipation rate. By numerical simulations, we compare the results by the non-Hermitian Hamiltonian approach and the Lindblad-form quantum master equation, which are not shown here. We find that the non-Hermitian Hamiltonian approach faithfully reproduce the key characteristics of the Lindblad-form quantum master equation approach although it may overestimate the decoherence.
Here, these three energy levels $d,~e,~m$ constitute $\Lambda$ configuration for the EIT. As shown in Appendix~\ref{AppendixA}, we can diagonalize  Hamiltonian~(\ref{eq:6}) as
\begin{eqnarray}
H_{B}'&=&E_{1}|E_{1}\rangle\langle E_{1}|+E_{2}|E_{2}\rangle\langle E_{2}|+E_{3}|E_{3}\rangle\langle E_{3}|,\label{eq:7}\\
|E_{i}\rangle&=&\frac{1}{N_{i}}\{[(E_{i}-\Omega_d-\omega_{1})(E_{i}-\Omega_d-\omega_{2})-\Omega_{c}^{2}]|d\rangle\nonumber\\
             & &+\Omega_{p}(E_{i}-\Omega_d-\omega_{2})|e\rangle+\Omega_{p}\Omega_{c}|m\rangle\},\label{eq:8}
\end{eqnarray}
where $|E_{i}\rangle's$ are the eigen states with eigen energies $E_{i}'s$, $\omega_1=\Delta_{e}-\Omega_{d}$, and $\omega_{2}=\Omega_{m}-\Omega_{d}$. The normalization $N_{i}'s$ constants are given by
\begin{eqnarray}
N_{i}^{2}&=&|(E_{i}-\Omega_d-\omega_{1})(E_{i}-\Omega_d-\omega_{2})-\Omega_{c}^{2}|^{2}\nonumber\\
         & &+|\Omega_{p}(E_{i}-\Omega_d-\omega_{2})|^{2}+|\Omega_{p}\Omega_{c}|^{2}.
\end{eqnarray}

By tuning $g_1$ and $g_2$ to satisfy $g_{1}/g_{2}=-\Omega_{c}/\Omega_{p}$, we have $|E_{1}\rangle=(g_{1}|d\rangle+g_{2}|m\rangle)/g$ with $g=\sqrt{g_1^2+g_2^2}$ according to Eq.~(\ref{eq:8}). 
The eigenvalues of the Hamiltonian are
\begin{align}
E_{1} &\simeq \frac{\Omega_{p}^{2}}{\Omega^{2}}\Omega_{m} + \left(1 - \frac{\Omega_{p}^{2}}{\Omega^{2}}\right)\Omega_{d}^{\prime} - i\frac{\Omega_{c}^{2}}{2\Omega^{2}}\kappa,\\
E_{2} &\simeq \Omega + \Omega^{\prime} - i\frac{\Omega_{p}^{2}}{4\Omega^{2}}\kappa,\\
E_{3} &\simeq -\Omega + \Omega^{\prime} - i\frac{\Omega_{p}^{2}}{4\Omega^{2}}\kappa,
\end{align}
where $\Omega^{2}=\Omega_{c}^{2}+\Omega_{p}^{2}$, $\Omega^{\prime}=\Delta_{e}+\Omega_{c}^{2}\Omega_{m}/\Omega^{2}+\left(1-\Omega_{c}^{2}/\Omega^{2}\right)\Omega_{d}^{\prime}$. We can see that when $\Omega_{c} \ll \Omega_{p}$, the dissipation rate of $|E_{1}\rangle$ is much smaller than  those of $|E_{2}\rangle$ and $|E_{3}\rangle$. Therefore, $|E_{1}\rangle$ is the dark state.
In the bases $|E_{j}\rangle~(j=1,2,3)$, Hamiltonian~(\ref{eq:5}) can be simplified as
\begin{eqnarray}
H_\textrm{eff}&=&E_{1}|E_{1}\rangle\langle E_{1}|+\sum_{k}(\omega_{k}a_{k}^{\dagger}a_{k}+Ja_{k}^{\dagger}|g\rangle\langle E_{1}|+\rm h.c.),\nonumber\\
\label{eq:9}
\end{eqnarray}
where $J=g/\sqrt{N}$. Notice that $|E_2\rangle$ and $|E_3\rangle$ are decoupled with the coupled cavities. In other words, we obtain a hybrid system of a two-level atom and a coupled-cavity array. 
In this two-level atom, the excited state is the weakly-dissipative state $|E_1\rangle$, while $|E_2\rangle$ and $|E_3\rangle$ do not interact with the cavity.

In our scheme for QB, we initially inject a photon into one of the cavities. Due to the coupling between the cavities, the photon will transfer between the cavities in the array. When the photon jumps the cavity where the QB is located, the QB will be charged by the photon because of the interaction between the cavity array and the QB. Since the total number of excitations is conserved due to Eq.~(\ref{eq:9}), in the single-excitation subspace, the wavefunction at time $t$ reads $|\psi(t)\rangle =u(t)|0,E_{1}\rangle+\sum_{k}\beta_{k}(t)a_{k}^{\dagger}|0,g\rangle$, where  $|0\rangle$ is the vacuum state of the coupled-cavity array, the initial condition is $u(0)=0$ and $\beta_k(0)=1/\sqrt{N}$. Here, $u(t)$ is the probability amplitude the of the cavities in the vacuum while the atom is in the dark state, and $\beta_k(t)$ is the probability amplitude of the state where there is a photon in the $k$th cavity mode while the atom is in the ground state. By the Schr\"{o}dinger equation, we can obtain
\begin{eqnarray}
i\dot{u}&=&E_{1}u+J\sum_{k}\beta_{k},\label{eq:10}\\
i\dot{\beta_{k}}&=&\omega_{k}\beta_{k}+Ju.\label{eq:11}
\end{eqnarray}
Equations~(\ref{eq:10},\ref{eq:11}) reveal that because of the interaction between the atom and the cavity, the energy of the photon can be transferred to the QB. Applying the Laplace transformation to Eqs.~(\ref{eq:10},\ref{eq:11}) yields
\begin{eqnarray}
\tilde{u}(p)&=&\frac{J}{\sqrt{N}}\frac{\sum_{k}\frac{1}{ip-\omega_{k}}}{p+iE_{1}+\sum_{k}\frac{J^{2}}{p+i\omega_{k}}}.\label{eq:12}
\end{eqnarray}
Although Eqs.~(\ref{eq:10},\ref{eq:11}) can only be solved numerically,
we can find approximate solution in the long-time limit by using the inverse Laplace transform on Eq.~(\ref{eq:12}). The singularities of Eq.~(\ref{eq:12}) are determined by
\begin{eqnarray}
p+i E_{1}+\sum_{k}\frac{J^{2}}{p+i\omega_{k}}=0.\label{eq:13}
\end{eqnarray}
Here, the branch cuts are defined by $p+i\omega_{k}=0$, with $p\in[ip_m, ip_M]$, where $p_m =-\omega_0-2\xi$ and $p_M =-\omega_0 + 2\xi$ are respectively the lowest and highest energy of the energy band of the coupled cavities. If $p\notin[-i(\omega_{0}+2\xi),-i(\omega_{0}-2\xi)]$, we can obtain
\begin{eqnarray}
p+i E_{1}+\frac{N}{2\pi\xi}\oint_{|z|=1}dz\frac{J^2}{z^{2}+Mz+1}=0,\label{eq:14}
\end{eqnarray}
where $M=(ip-\omega_{0})/\xi$. When $ip>\omega_0 + 2\xi$, i.e., $M>2$, the solution is denoted as $p_1$, while for $ip<\omega_0 - 2\xi$, i.e., $M<-2$, the solution is denoted as $p_2$. The detailed calculation is shown in Appendix~\ref{AppendixB}.

It is worthy noticing that the solutions of Eq.~(\ref{eq:14}) are the eigen energy of Hamiltonian~(\ref{eq:9}). By Schr\"{o}dinger equation $H|\Phi\rangle=E|\Phi\rangle$, we can obtain
\begin{eqnarray}
 E_1+\sum_{k}\frac{ J^{2}}{E-\omega_{k}}&=& E.\label{eq:15}
\end{eqnarray}
Let's define $E=ip$. Obviously, Eq.~(\ref{eq:13}) and Eq.~(\ref{eq:15}) are equivalent. Since Eq.~(\ref{eq:14}) determines the eigen energies of the hybrid system, it governs the dynamical evolution of the QB. In other words, it is the spectral characteristics of the whole system that determine the dynamics of the QB. At the same time, we know that the branch cut with $p\in[ip_m, ip_M]$ corresponds to an energy band, while $p_1$ and $p_2$ correspond to the two bound states. By substituting the branch cut and the two singularities of Eq.~(\ref{eq:14}) into the inverse Laplace transformation $u(t)=\int_{\sigma-i\infty}^{\sigma+i\infty}dp\tilde{u}(p)e^{pt}/2\pi i$ of Eq.~(\ref{eq:12}), we have
\begin{eqnarray}
u(t)\!\!&=&\!\!B_{1}Q(p_{1})e^{p_{1}t}+B_{2}Q(p_{2})e^{p_{2}t}+\int_{-2\xi}^{2\xi}C(x)e^{i(x-\omega_{0})t}dx,\nonumber\\
\label{eq:16}
\end{eqnarray}
where
\begin{eqnarray}
Q(p_{1})\!\!\!&=&\!\!\!\frac{g}{\xi\sqrt{M^{2}-4}}=-Q(p_{2}),\label{eq:17}\\
B_j\!\!\!&=&\!\!\!\frac{(ip_{j}-\omega_{0})^{2}-4\xi^{2}}{(ip_{j}-\omega_{0})^{2}-4\xi^{2}+(ip_{j}-E_{1})(ip_{j}-\omega_{0})},
   \label{eq:19}\\
C(x)\!\!\!&=&\!\!\!-\frac{1}{\pi}\frac{\frac{g}{\sqrt{4\xi^{2}-(x)^{2}}}[(x-\omega_{0})+E_{1}]}{(E_{1}-\omega_{0}+x)^{2}+\frac{J^{2}N^{2}}{4\xi^{2}-(x)^{2}}}e^{i(x-\omega_{0})t}.\label{eq:20}
\end{eqnarray}
Because the energy band $p\in[ip_m, ip_M]$ of the system leads to the occurrence of the branch cut, the contribution of the branch cut tends to zero for a long time due to out-of-phase interference \cite{Ji2022PRB,Liu2016PRA}. We can obtain the probability  $P_{E_1}=|u(\infty)|^2$ in the long-time limit as
\begin{eqnarray}
P_{E_1}  \equiv  Q(p_{1})^2[B_{1}^2  + B_{2}^2  - 2 B_{1} B_{2}  \cos(\phi t)],
\label{eq:21}
\end{eqnarray}
where $\phi=ip_1-ip_2$. When the dissipation $\kappa$ of the system is absent, $\phi$ is obviously real. As a result, there is coherent energy exchange between the QB and the charger the energy as manifested by the Rabi-like oscillation due to existence of the two bound states. However, when there is only one bound state, $P_{E_1}  \equiv  Q(p_{1})^2B_{1}^2$ and thus the Rabi-like oscillations disappear. Because of the second law of thermodynamics, not all of the QB's energy can be converted to work \cite{Monsel2020PRL}. Thus, in the rotating frame we can use the ergotropy \cite{Francica2020PRL,Cakmak2020PRE} to evaluate the amount of extractable work of QB via
\begin{eqnarray}
\mathcal{W}(t)=\mathrm{Tr}[\rho'_B(t)H'_B]-\mathrm{Tr}[\tilde{\rho}'_B(t)H'_B],\label{eq:22}
\end{eqnarray}
where $\rho'_B(t)=\mathrm{Tr}_c(|\psi'(t)\rangle\langle\psi'(t)|)$ is the reduced density matrix of the QB, $\tilde{\rho}'_B(t)=\sum_{k}{r_k(t)|\epsilon_k}\rangle\langle\epsilon_k|$ is the passive state of the QB. $r_k(t)$'s are the eigenvalues of $\rho'_B(t)$ in descending order, and $|\epsilon_k\rangle$'s are the eigenstates of $H'_B$ corresponding to the eigenvalues $\epsilon_k$'s in ascending order. The rotating frame is equivalent to the interaction picture, which aims at eliminating the time-dependence of the Hamiltonian and thus enables simplification of the calculation and even obtain the analytical solution. That is because the transformation between the different frames should not modify the results of any physical quantity but simplify the calculation.


During the charging process, the photon is initially injected into the cavity where the atom is located.
By carefully tuning the parameters such that $g_1/g_2 = -\Omega_c/\Omega_p$, the dark state and the two bright states are formed and only the dark state is coupled to the cavity modes, while the bright states become decoupled.
Consequently, the photon interacts selectively with the dark state $|E_1\rangle$, enabling coherent energy transfer from the cavity array to the QB, due to the nature of the EIT.

\section{Numerical Simulation and Discussion}
\label{sec:Discussion}

\begin{figure}[htbp]
\centering
\includegraphics[bb=00 0 380 300,width=8.5cm]
{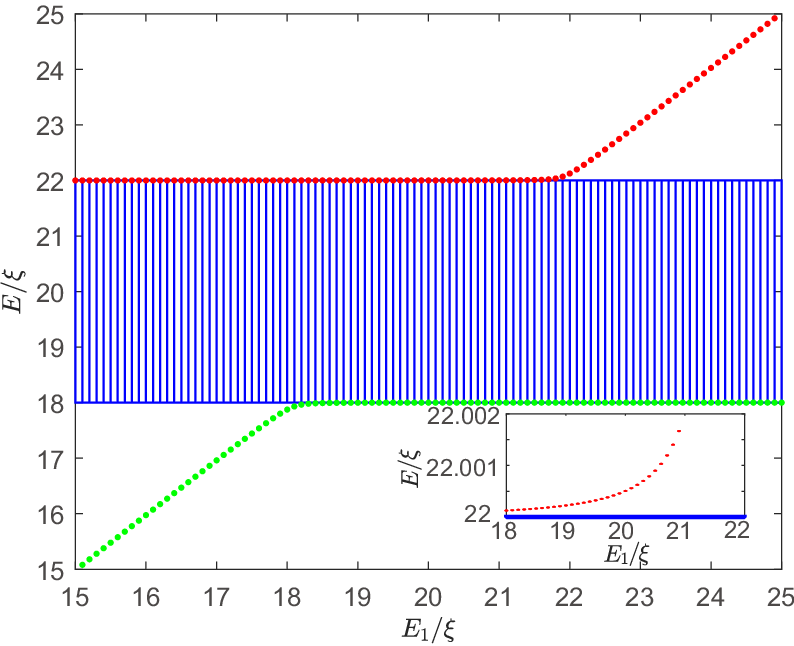}
\caption{The dependence of the energy of the bound states on the energy band and the energy $E_1$ of the QB. The blue solid line is the energy band $\omega_k=\omega_0-2\xi\cos (k)$ of the coupled-cavity array. The energies of the bound states denoted by the red and green dots are obtained by Eq.~(\ref{eq:15}). The parameters used in the calculation are $\omega_0 = 20\xi$, $g = 0.3\xi$. }\label{fig:2}
\end{figure}

In our scheme, the energy is stored in the composite system of the QB and the charger as shown by the Rabi-like oscillations in Eq.~(\ref{eq:21}). This implies that the number of the bound states in the open quantum system is important. Therefore, in Fig.~\ref{fig:2}, we demonstrate the energy of the bound states and the energy band $\omega_k=\omega_0-2 \xi \cos(k)$ of the total system. It is shown that when the energy $E_1$ of the QB is within the energy band, there are two bound states. However, when $E_1$ is outside of the energy band, only one bound state is present. In other words, we can effectively modify the number of the bound state in the composite system by tuning the energy of the QB.

\begin{figure}[htbp]
\centering
\includegraphics[bb=00 0 380 550,width=8.3cm]
{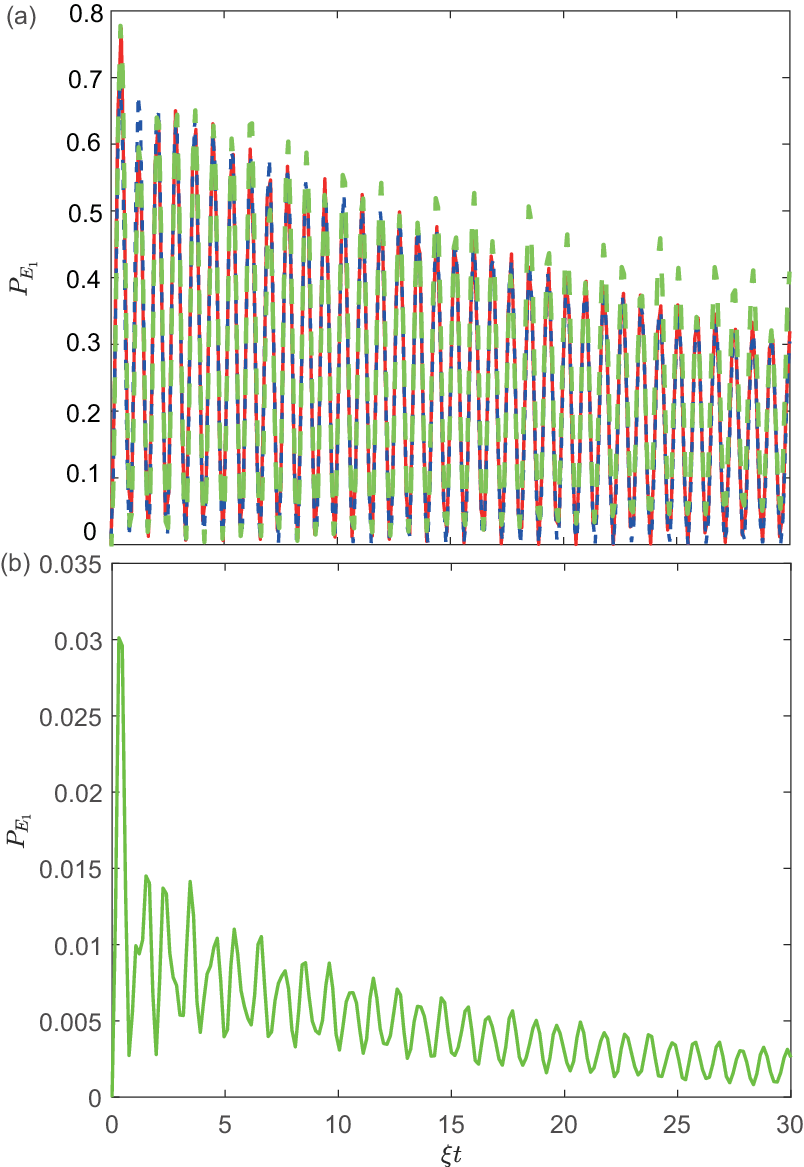}
\caption{Probability of the dark state $|E_1\rangle$ vs time $t$. (a) When $E_1$ is in the energy band of the coupled cavities, the analytical results obtained by Eq.~(\ref{eq:21}) are denoted as the blue dots while the numerical results are denoted by the red solid line. In addition, the green dashed line is calculated using the Lindblad-form quantum master equation. The parameters are as follows, $E_1/\xi=35.327-0.032i$, $\omega_0 = 100\xi/3$, $\Omega_d=106\xi/3$, $\Omega_e=50\xi$, $\Omega_m=106\xi/3$, $\kappa=20\xi/3$, $\Omega_p=50\xi/3$, $\Omega_c=5\xi/3$, and the number of cavities $N=253$. (b) When $E_1$ is outside of the energy band of the coupled cavities, the numerical results denoted by green solid line are obtained by $E_1/\xi=66.6553 - 0.0287i$, $\omega_0 = 100\xi/3$, $\Omega_d=200\xi/3$, $\Omega_e=100\xi$, $\Omega_m=200\xi/3$, $\kappa=20\xi/3$, $\Omega_p=50\xi/3$, $\Omega_c=5\xi/3$, and $N=253$.}\label{fig:3}
\end{figure}

In Fig.~\ref{fig:3} we investigate the effects of the number of the bound states in the composite system on the probability of the dark state of the QB. Therein, we compare the analytical solutions by Eq.~(\ref{eq:21}) and the numerical simulations. It is shown that the analytical solution coincides with the numerical simulation except for the short-time regime, since we neglect the contribution from the branch cut.
In addition to the coherent oscillations, the probability decays due to the dissipation of the QB. Similarly, we also compare the analytical solution and numerical simulation with the Lindblad-form quantum master equation method \cite{breuer2002}. We show that our results are in good agreement with those by the quantum master equation. In Fig.~\ref{fig:3}(a), we observe a significant Rabi-like coherent oscillation between the atoms and the cavity array. This oscillations reflect the formation of two bound states in the system of atom-cavity-array, i.e., the energy is repeatedly exchanged between the atom and the local cavities without propagating to the cavities far away from the atom. As a result, the energy is localized in the vicinity of the atom, creating an effective ``protection zone" that significantly inhibits energy losses to the cavity array. On the contrary, when the energy of the dark state of the QB is outside of the energy band in Fig.~\ref{fig:3} (b), we find that the probability of the QB being in the dark state is very low, and the oscillation over time is irregular, which is bad for the storage of energy.


\begin{figure}[htbp]
\centering
\includegraphics[bb=20 0 400 300,width=8.8cm]
{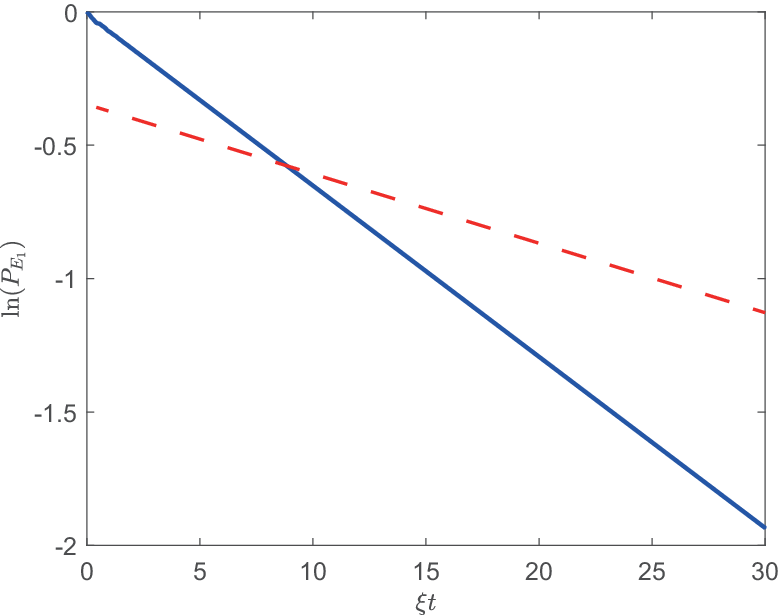}
\caption{The envelope of the probability of the dark state with/without two bound states vs time. The red dashed line denotes the case when there are two bound states. We use the function $\ln(P_{E_1})=-0.35309-2.6487\times10^{-2}\xi t$ to fit the data with the linear correlation coefficient $|r|$ = 0.99314. The blue solid line shows the case when there is no bound state with the corresponding linear correlation coefficient $|r|$ = 1.0000. Initially, the atom is in the state $\ket{m}$. The parameters are as follows, i.e., $E_1/\xi=35.327-0.032i$, $\Omega_d=106\xi/3$, $\Omega_e=50\xi$, $\Omega_m=106\xi/3$, $\kappa=20\xi/3$, $\Omega_p=50\xi/3$, $\Omega_c=5\xi/3$. }\label{fig:4}
\end{figure}

In realistic systems, due to the couplings to the environment, there would exist quantum decoherence. In Fig.~\ref{fig:4}, we plot the envelope of the probability of which the atom is in the dark state in order to explore the effects of the bound states on the energy storage. At the very beginning, the envelope changes somewhat chaotically, which is not shown here. However, after some time it tends to decay exponentially. By numerical fitting as shown by the red dashed line, we obtain an effective decay rate $\kappa'=2.6487\times10^{-2}\xi \ll \kappa=20\xi/3$ when there are two bound states. If we neglect the effect of the cavity array, i.e., exclude the contribution from the bound states, the decay rate of $P_{E_1}(t)$ is given by $\gamma$. By numerical fitting as shown by the blue solid line, we obtain the decay rate $\gamma = 0.0645\,\xi \approx 2\,\kappa'$. It can be seen that thanks to the presence of the bound states, the decay rate is further reduced as compared to the case when only the EIT is present. Also, this implies that due to the EIT and the bound states, the dissipation has been significantly inhibited and thus suppress the degradation of the QB.

\begin{figure}[htbp]
\centering
\includegraphics[bb=10 0 400 300,width=8.8cm]
{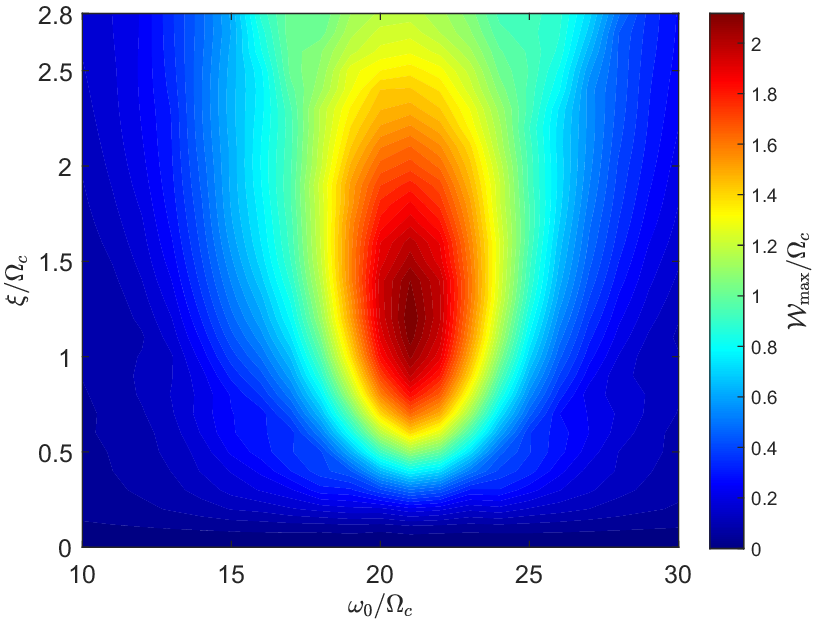}
\caption{The maximum ergotropy $\mathcal{W}_{\max}$ during the duration $[0,t_{\rm max}=15/\Omega_c]$ vs the coupling strength $\xi$ and the frequency $\omega_0$ of the cavity. The other parameters are $\Omega_d=21.2\Omega_c$, $\Omega_e=30\Omega_c$, $\Omega_m=21.2\Omega_c$, $\kappa=4\Omega_c$, $\Omega_p=10\Omega_c$, and $N=253$.}\label{fig:5}
\end{figure}

Another important physical parameter to assess the QB is the extractable energy, i.e., the ergotropy \cite{Monsel2020PRL,Francica2020PRL} in Eq.~(\ref{eq:22}). Since the coupled-cavity array acts as a QB charger, and we shall investigate the relationship between its ability to charge the QB and the parameters. Here, we initially inject the photon into the adjacent cavity of the $0th$ cavity, and explore the dynamics of the charging of QB. The relationship between the ergotropy $\mathcal{W}$ and the frequency $\omega_0$ of the cavity and the coupling strength $\xi$ between two cavities is shown in Fig.~\ref{fig:5}. For the present set of parameters, we find that no matter how $\xi$ changes, there is a maximum value of $\mathcal{W}$ when $\omega_0$ is equal to the energy $E_1$ of the dark state. This implies that both the useful work and the efficiency of the QB are the largest when the cavity and the QB are in resonance.

\begin{figure}[htbp]
\centering
\includegraphics[bb=20 0 400 300,width=8.8cm]
{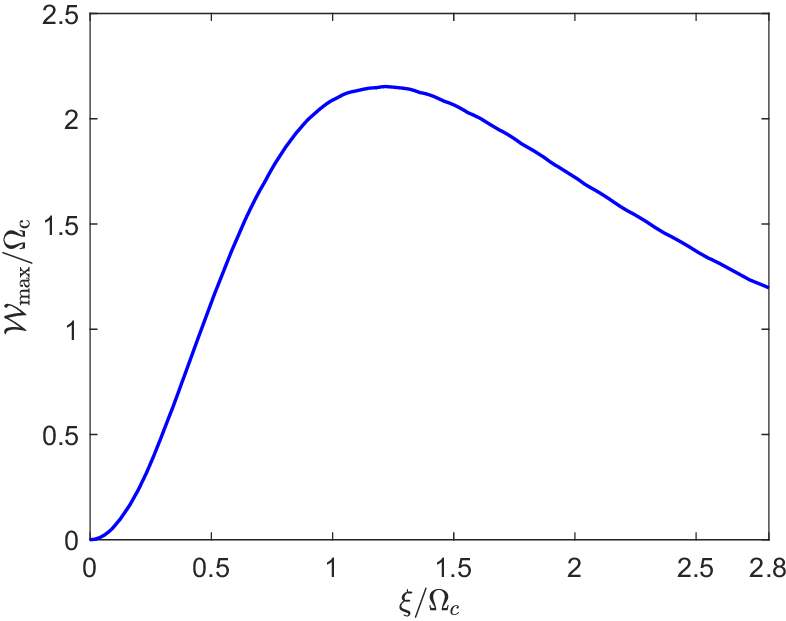}
\caption{The dependence of the maximum ergotropy $\mathcal{W}_{\rm max}$ in the period $[0, t_{\rm max}=15/\Omega_c]$ on the coupling between two cavities $\xi$ when $\omega_0=\textrm{Re}(E_1)$. The other parameters $\omega_0 = 21.196\Omega_c$, $\Omega_d=21.2\Omega_c$, $\Omega_e=30\Omega_c$, $\Omega_m=21.2\Omega_c$, $\kappa=2\Omega_c$, $\Omega_p=10\Omega_c$, and $N=253$. }\label{fig:6}
\end{figure}

\begin{figure}[htbp]
\centering
\includegraphics[bb=20 0 400 300,width=8.8cm]
{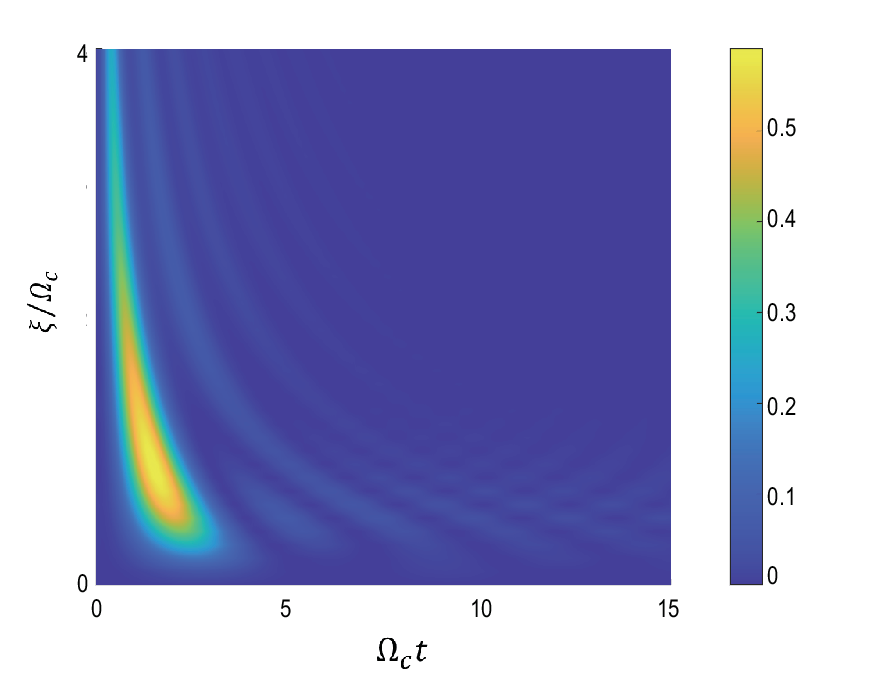}
\caption{The average charging power $P/\Omega_{c}^2$ as a function of the coupling strength $\xi$ and time. The parameters are $\omega_0 = 21.196\Omega_c$, $\Omega_d=21.2\Omega_c$, $\Omega_e=30\Omega_c$, $\Omega_m=21.2\Omega_c$, $\kappa=4\Omega_c$, $\Omega_p=10\Omega_c$, and $N=253$.}\label{fig:10}
\end{figure}
In Fig.~\ref{fig:6}, we explore the relation between the maximum ergotropy $\mathcal{W}_{\rm max}$ of the QB and the coupling strength $\xi$ under the resonance condition with $E_1=(21.196 - 0.019 i)\Omega_c$. We find that for small $\xi$, $\mathcal{W}_{\rm max}$ increases along with the rise of $\xi$, because the coupling between the cavities helps deliver the photon to the cavity where the atom is located. However, after it reaches the maximum around $\xi=1.22\Omega_c$, $\mathcal{W}_{\rm max}$ experiences an decrease when further increasing the coupling strength. This suggests that strong coupling between the cavities can prompts the delivered photon away from the cavity with the atom and thus suppresses the ergotropy. In summary, the coupling between the two cavities can improve the efficiency of the QB assisted by the resonance between the QB and the cavity.

Finally, we shall discuss the effects of charging time and coupling strength $\xi$ on the average charging power, which is defined as $P(t)=W(t)/t$ \cite{Dario2018PRL}. The average charging power is critical for QBs as it helps determine the optimal time to turn off the charging. When the average charging power is high, it indicates that the QB is charging rapidly. On the contrary, when the average charging power is low, it suggests that the change in the QB's ergotropy is slow. As shown in Fig.~\ref{fig:10}, we can observe that for a given coupling strength $\xi$, the charging power increases quickly and then decreases, eventually stabilizing at a relatively-low steady state. By varying $\xi$, we find that there is an optimal value of $\xi$ for maximizing the average charging power. A larger $\xi$ does not necessarily lead to better performance in the average charging power, and this insight can guide us in optimizing the QB in experiments.

In the above analysis of the effects of the photon frequency $\omega_0$ and the coupling strength $\xi$ on the ergotropy and the average charging power of the QB, our purpose is to identify the optimal experimental parameters. The QB can be optimized in advance according to the requirements of the load, thereby fixing its parameters. During the charging process, these parameters remain unchanged.

\section{Conclusion}\label{sec:Conclusion}

In this paper, we propose a scheme for charging QB by coupling QB with cavity array. With the EIT, we obtain the QB of an effective two-level system including the dark state and the ground state. When the energy of the dark state is within the energy band of the cavity array, there are two bound states. We show that the energies of the bound states determine the distribution dynamics of QB in the excited state, resulting in Rabi-like oscillations. Under the influence of the bound states and the dark state, we find that the dissipation of the environment is strongly suppressed. When the energy of the dark state is resonant with the bare frequency of the cavity, the extractable work of the QB reaches its maximum. In addition, we demonstrate that the coupling strength of the two cavities maximizes the extractable work of the QB. Our scheme extends the QB's lifetime and avoids the reduction in charging efficiency by suppressing environmental decoherence, providing an avenue for the efficient QB charging.

\begin{acknowledgments}

 This work is supported by Innovation Program for Quantum Science and Technology under Grant No.~2023ZD0300200, the National Natural Science Foundation of China under Grant No.~62461160263, Beijing Natural Science Foundation under Grant No.~1202017, and Beijing Normal University under Grant No.~2022129.

\end{acknowledgments}

\appendix

\section{Diagonalization of Atomic Hamiltonian}
\label{AppendixA}

In this section, we will obtain the eigenvalues and eigenstates of atomic Hamiltonian by the perturbation theory.

After the Fourier transformation, the total Hamiltonian is given by
\begin{eqnarray}
H&=&\sum_{k}\omega_{k}a_{k}^{\dagger}a_{k}+\frac{1}{\sqrt{N}}\sum_{k}g_{1}(a_{k}^{\dagger}|g\rangle\langle d|+a_{k}|d\rangle\langle g|)\nonumber\\
& & +\frac{1}{\sqrt{N}}\sum_{k}g_{2}(a_{k}^{\dagger}|g\rangle\langle m|+a_{k}|m\rangle\langle g|))+H_0,\label{eq:A1}
\end{eqnarray}
where
\begin{eqnarray}
H_B&=&\Omega_{d}|d\rangle\langle d|+\Omega_{e}|e\rangle\langle e|+\Omega_{m}|m\rangle\langle m|\nonumber\\
& &+\Omega_{p}e^{-i\omega_{p}t}|e\rangle\langle d|+\Omega_{c}e^{-i\omega_{c}t}|e\rangle\langle m|+\rm h.c.,\label{eq:A2}
\end{eqnarray}
with $\omega_k=\omega_0-2\xi \cos(k)$, $\Omega_{d}=\Omega'_{d}-i\kappa/2$. In the rotating frame defined by $U_R=\exp(-i\omega_{e}t|e\rangle\langle e|)$, the Hamiltonian (\ref{eq:A2}) is rewritten as
\begin{eqnarray}
H_B'&=&U_{R}^{\dagger}H_0U_{R}-iU_{R}^{\dagger}\dot{U_{R}}\nonumber\\
    &=&\Omega_{d}|d\rangle\langle d|+(\Omega_{e}-\omega_{e})|e\rangle\langle e|+\Omega_{m}|m\rangle\langle m|\nonumber\\
    & &+\Omega_{p}e^{-i(\omega_{p}-\omega_{e})t}|e\rangle\langle d|+\Omega_{c}e^{-i(\omega_{c}-\omega_{e})t}|e\rangle\langle m|+\rm h.c.\nonumber\\
    \label{eq:A4}
\end{eqnarray}
Assuming $\omega_{p}=\omega_{e}=\omega_{c}$, we obtain
\begin{eqnarray}
H_B'&=&\Delta_{e}|e\rangle\langle e|+\Omega_{d}|d\rangle\langle d|+\Omega_{m}|m\rangle\langle m|\nonumber\\
& &+\Omega_{p}|e\rangle\langle d|+\Omega_{c}|e\rangle\langle m|+\rm h.c.,\label{eq:A5}
\end{eqnarray}
with $\Delta_{e}=\Omega_{e}-\omega_{e}$. Defining $E=y+\Omega_{d}$, the eigenvalue $E$ of Hamiltonian (\ref{eq:A5}) is determined by
\begin{eqnarray}
\begin{vmatrix}-y & \Omega_{p}\\
\Omega_{p} & \omega_{1}-y & \Omega_{c}\\
 & \Omega_{c} & \omega_{2}-y
\end{vmatrix}&=&0,\label{eq:A7}
\end{eqnarray}
with $\omega_{1}=\Delta_{e}-\Omega_{d},\omega_{2}=\Omega_{m}-\Omega_{d}$, which is equivalent to
\begin{eqnarray}
y^{3}-y^{2}(\omega_{1}+\omega_{2})+y(\omega_{1}\omega_{2}-\Omega_{c}^{2}-\Omega_{p}^{2})\!\!=\!\!-\Omega_{p}^{2}\omega_{2}.\label{eq:A8}
\end{eqnarray}
If we assume $\omega_2$ is small, to the first order of $\omega_2$, the solutions can be written as $y_{j}=y_{0j}+A_{j}\omega_{2}~(j=1,2,3)$. By the perturbation theory, we can approximate Eq.~(\ref{eq:A8}) to the zeroth-order as
\begin{eqnarray}
y^{3}-y^{2}(\omega_{1}+\omega_{2})+y(\omega_{1}\omega_{2}-\Omega_{c}^{2}-\Omega_{p}^{2})&=&0.
\end{eqnarray}
Thus, we can obtain
\begin{eqnarray}
y_{01}&=&0,\\
y_{02}&=&\omega_{+},\\
y_{03}&=&\omega_{-},
\end{eqnarray}
with $\Omega^{2}=\Omega_{c}^{2}+\Omega_{p}^{2}$, $\omega_{\pm}=[(\omega_{1}+\omega_{2})\pm\sqrt{(\omega_{1}-\omega_{2})^{2}+4\Omega^{2}}]/2$. If $\omega_{1},\omega_{2}\ll\Omega$, we can approximate $\omega_{\pm}$ as
\begin{eqnarray}
\omega_{\pm}&\simeq&\frac{1}{2}\left\{\omega_{1}+\omega_{2}\pm2\Omega\left[1+\frac{(\omega_{1}-\omega_{2})^{2}}{4\Omega^{2}}\right]\right\}\nonumber\\
&\simeq&\frac{1}{2}[\omega_{1}+\omega_{2}\pm2\Omega].\label{eq:A9}
\end{eqnarray}
Substituting the zeroth-order solutions $y_{0j}$'s into Eq.~(\ref{eq:A8}) yields
\begin{eqnarray}
A_{1}&=&-\frac{\Omega_{p}^{2}}{\omega_{+}\omega_{-}},\\
A_{2}&=&-\frac{\Omega_{p}^{2}}{\omega_{+}(\omega_{+}-\omega_{-})},\\
A_{3}&=&\frac{\Omega_{p}^{2}}{\omega_{-}(\omega_{+}-\omega_{-})}.
\end{eqnarray}
In conclusion, we could obtain the eigenvalues of Hamiltonian $H_0'$ as
\begin{eqnarray}
E_{1}&=&-\frac{\Omega_{p}^{2}}{1/4*[(\omega_{1}+\omega_{2})^{2}-4\Omega^{2}]}\omega_{2}+\Omega_{d}\nonumber\\
     &\simeq&\frac{\Omega_{p}^{2}}{\Omega^{2}}\omega_{2}+\Omega_{d},\\
E_{2}&=&\frac{1}{2}[\omega_{1}+\omega_{2}+2\Omega]-\frac{\Omega_{p}^{2}}{\omega_{+}(\omega_{+}-\omega_{-})}\omega_{2}\nonumber\\
     &\simeq&\Omega+\frac{1}{2}(\omega_{1}+\frac{\Omega_{c}^{2}}{\Omega^{2}}\omega_{2})+\Omega_{d},\\
E_{3}&=&\frac{1}{2}[\omega_{1}+\omega_{2}-2\Omega]+\frac{\Omega_{p}^{2}}{\omega_{+}(\omega_{+}-\omega_{-})}\omega_{2}\nonumber\\
     &\simeq&-\Omega+\frac{1}{2}(\omega_{1}+\frac{\Omega_{c}^{2}}{\Omega^{2}}\omega_{2})+\Omega_{d}.
\end{eqnarray}
Correspondingly, the eigenstates are respectively
\begin{eqnarray}
|E_{i}\rangle&=&\frac{1}{N_{i}}\{[(E_{i}-\Omega_d-\omega_{1})(E_{i}-\Omega_d-\omega_{2})-\Omega_{c}^{2}]|d\rangle\nonumber\\
             & &+\Omega_{p}(E_{i}-\Omega_d-\omega_{2})|e\rangle+\Omega_{p}\Omega_{c}|m\rangle\},\
\end{eqnarray}
where the normalization constants are given by
\begin{eqnarray}
N_{i}^{2}&=&|(E_{i}-\Omega_d-\omega_{1})(E_{i}-\Omega_d-\omega_{2})-\Omega_{c}^{2}|^{2}\nonumber\\
         & &+|\Omega_{p}(E_{i}-\Omega_d-\omega_{2})|^{2}+|\Omega_{p}\Omega_{c}|^{2}.
\end{eqnarray}
To the zeroth-order of $\omega_{j}$'s,
we can simplify the eigenstates as
\begin{eqnarray}
|E_{1}\rangle&=&\frac{1}{N_{1}}(-\Omega_{c}^{2}|d\rangle+\Omega_{p}\Omega_{c}|m\rangle),\label{eq:A25}\\
|E_{2}\rangle&=&\frac{1}{N_{2}}(\Omega_{p}^{2}|d\rangle+\Omega_{p}\Omega|e\rangle+\Omega_{p}\Omega_{c}|m\rangle),\label{eq:A26}\\
|E_{3}\rangle&=&\frac{1}{N_{3}}(\Omega_{p}^{2}|d\rangle-\Omega_{p}\Omega|e\rangle+\Omega_{p}\Omega_{c}|m\rangle),\label{eq:A27}
\end{eqnarray}
with $N_{1}=\Omega\Omega_{c}$, $N_{2}=\sqrt{2}\Omega\Omega_{p}$, $N_{3}=\sqrt{2}\Omega\Omega_{p}$. Here, $|E_{1}\rangle$ is the dark state, while $|E_{2}\rangle$ and $|E_{3}\rangle$ are the bright states.
Furthermore, the bare atomic states can rewritten in terms of the eigenstates as
\begin{eqnarray}
|d\rangle&=&\frac{1}{2\Omega^{2}}(N_{2}|E_{2}\rangle+N_{3}|E_{3}\rangle-2N_{1}|E_{1}\rangle),\\
|e\rangle&=&\frac{1}{2\Omega_{p}\Omega}(N_{2}|E_{2}\rangle-N_{3}|E_{3}\rangle),\\
|m\rangle&=&\frac{1}{2\Omega_{p}\Omega_{c}}(N_{2}|E_{2}\rangle+N_{3}|E_{3}\rangle+2N_{1}|E_{1}\rangle).
\end{eqnarray}

\section{Evolution of the Dark State}
\label{AppendixB}


In the rotating frame defined by $U_R=\exp(-i\omega_{e}t|e\rangle\langle e|)$, the Hamiltonian of the hybrid system including the four-level atoms and the cavity array can be transformed as
\begin{eqnarray}
H'&=&\sum_{k}\omega_{k}a_{k}^{\dagger}a_{k}+\frac{1}{\sqrt{N}}[\sum_{k}g_{1}(a_{k}^{\dagger}|g\rangle\langle d|+a_{k}|d\rangle\langle g|)\nonumber\\
 & &+\sum_{k}g_{2}(a_{k}^{\dagger}|g\rangle\langle m|+a_{k}|m\rangle\langle g|)]+H_{0}'.\label{eq:B2}
\end{eqnarray}
Assuming $g_{1}/g_{2}=-\Omega_{c}/\Omega_{p}$, we can obtain $|E_{1}\rangle=(g_{1}|d\rangle+g_{2}|m\rangle)/g$ with $g=\sqrt{g_1^2+g_2^2}$. In the bases $|E_{1}\rangle~(j=1,2,3)$, Hamiltonian~(\ref{eq:B2}) can be simplified as
\begin{eqnarray}
H_\textrm{eff}&=&\sum_{k}\omega_{k}a_{k}^{\dagger}a_{k}+E_{1}|E_{1}\rangle\langle E_{1}|\nonumber\\
&&+J\sum_{k}(a_{k}^{\dagger}|g\rangle\langle E_{1}|+a_{k}|E_{1}\rangle\langle g|),\label{eq:B3}
\end{eqnarray}
with $J=g/\sqrt{N}$.
Note that the bright states $|E_2\rangle$, $|E_3\rangle$ are decoupled from the cavity modes in Eq.~(\ref{eq:B3}). In the single-excitation subspace, the state of the total system reads
\begin{eqnarray}
|\psi(t)\rangle =u(t)|0,E_{1}\rangle+\sum_{k}\beta_{k}(t)a_{k}^{\dagger}|0,g\rangle.\label{eq:B4}
\end{eqnarray}
Here, $|0,E_{1}\rangle$ is the cavities in the vacuum while the atom is in the dark state. $a_{k}^{\dagger}|0,g\rangle$ is the state where there is a photon in the $k$th cavity mode while the atom is in the ground state.
Using the Schr\"{o}dinger equation for Eqs.~(\ref{eq:B3},\ref{eq:B4}), we can obtain
\begin{eqnarray}
i\dot{u}&=&E_{1}u+J\sum_{k}\beta_{k},\label{eq:B5}\\
i\dot{\beta_{k}}&=&\omega_{k}\beta_{k}+Ju,\label{eq:B6}
\end{eqnarray}
where $u(0) = 0$, $\beta_k(0) = 1/\sqrt{N}$ are the initial condition of the system. Next, we apply a Laplace transformation to Eqs.~(\ref{eq:B5},\ref{eq:B6}) to obtain
\begin{eqnarray}
i[p\tilde{u}(p)-u(0)]&=&E_{1}\tilde{u}(p)+J\sum_{k}\tilde{\beta_{k}}(p),\label{eq:B7}\\
i[p\tilde{\beta_{k}}(p)-\beta_{k}(0)]&=&\omega_{k}\tilde{\beta_{k}}(p)+J\tilde{u}(p).\label{eq:B8}
\end{eqnarray}
With a bit of algebra, we can calculate the probability amplitude of $|0,E_{1}\rangle$ in $p$-space as follows
\begin{eqnarray}
\tilde{u}(p)&=&\frac{J}{\sqrt{N}}\frac{\sum_{k}\frac{1}{ip-\omega_{k}}}{p+iE_{1}+\sum_{k}\frac{J^{2}}{p+i\omega_{k}}}.\label{eq:B9}
\end{eqnarray}
The probability amplitude of $|0,E_{1}\rangle$ in the time domain can be obtained from the inverse Laplace transform as
\begin{eqnarray}
u(t)&=&\frac{1}{2\pi i}\int_{\sigma-i\infty}^{\sigma+i\infty}dp\tilde{u}(p)e^{pt}\nonumber\\
&=&\sum_{j}\textrm{res}\left[\tilde{u}(p_{j})e^{p_{j}t}\right]-\frac{1}{2\pi i}\left[\int_{C_{R}}dp\tilde{u}(p)e^{pt}\right.\nonumber\\
      &   &\left.+\int_{l_{1}}dp\tilde{u}(p)e^{pt}+\int_{l_{2}}dp\tilde{u}(p)e^{pt}\right].\label{eq:B11}
\end{eqnarray}
Here, $C_R$ is the large semi-circle in infinity, $p_j 's$ and $l_j 's$ $(j = 1, 2)$ are the two singularities and two branch cuts respectively, which are defined from
\begin{eqnarray}
p+iE_{1}+\sum_{k}\frac{J^{2}}{p+i\omega_{k}}&=&0.\label{eq:B12}
\end{eqnarray}
Following Eq.~(\ref{eq:B12}), the branch cuts are defined by $p + i\omega_k = 0$, with $p\in[ip_m, ip_M]$, where $p_m =-\omega_0-2\xi$,
$p_M =-\omega_0 + 2\xi$. Please note
\begin{eqnarray}
\sum_{k}\frac{1}{p+i\omega_{k}}&=&\frac{N}{2\pi}\int dk\frac{1}{p+i(\omega_{0}-2\xi \cos(k))}\nonumber\\
                               &=&\frac{N}{2\pi}\int_{-\pi}^{\pi}dk\frac{1}{p+i(\omega_{0}-\xi(e^{ik}+e^{-ik}))}\nonumber\\
                               &=&\frac{N}{2\pi}\oint_{|z|=1}dz\frac{1}{p+i(\omega_{0}-\xi(z+z^{-1}))}\frac{1}{iz}\nonumber\\
                               &=&\frac{N}{2\pi\xi}\oint_{|z|=1}dz\frac{1}{z^{2}+\frac{(ip-\omega_{0})z}{\xi}+1}\nonumber\\
                               &=&\frac{N}{2\pi\xi}\oint_{|z|=1}dz\frac{1}{z^{2}+Mz+1},\label{eq:B13}
\end{eqnarray}
where $M=(ip-\omega_{0})/\xi$, $z_{\pm}=[-M\pm\sqrt{M^{2}-4}]/2$ are the two singularities. Thus, if $p\notin[-i(\omega_{0}+2\xi),-i(\omega_{0}-2\xi)]$, we have $M>2$ or $M<-2$. For the former case, since $ip>\omega_{0}+2\xi$, i.e., $-1<z_{+}<0$, $z_{-}<-1$, we have
\begin{eqnarray}
&&\frac{N}{2\pi\xi}\oint_{|z|=1}dz\frac{1}{z^{2}+Mz+1}\nonumber\\
&=&\frac{N}{2\pi\xi}\oint_{|z|=1}dz\frac{1}{(z-z_{+})(z-z_{-})}\nonumber\\
&=&\frac{iN}{\xi\sqrt{M^{2}-4}}.\label{eq:B14}
\end{eqnarray}
In order to obtain the singularity, we substitute Eq.~(\ref{eq:B14}) into Eq.~(\ref{eq:B12}), and thus we have
\begin{eqnarray}
p+iE_{1}+J^{2}\frac{iN}{\xi\sqrt{M^{2}-4}}&=&f_{1}(p).\label{eq:B15}
\end{eqnarray}
Letting $f_1(p)=0$, we can obtain the singularity $p_1$ with $ip_{1}>\omega_{0}+2\xi$. Similarly, for the latter case, we can also obtain the second singularity $p_2$ with $ip_2<\omega_{0}-2\xi$ by
\begin{eqnarray}
p+iE_{1}-J^{2}\frac{iN}{\xi\sqrt{M^{2}-4}}&=&f_{2}(p).\label{eq:B16}
\end{eqnarray}

In the above calculations, we can obtain the contribution from the singularity in Eq.~(\ref{eq:B9}).
Hereafter, we will obtain the contribution from the branch cuts in Eq.~(\ref{eq:B9}). Because of Jordan Lemma, we can know $\int_{C_{R}}dp\tilde{u}(p)e^{pt}=0$. So far as the contribution from the branch cuts, that is $p\in[ip_m, ip_M]$, we can obtain
\begin{eqnarray}
&&\int_{l_{1}}dp\tilde{u}(p)e^{pt}+\int_{l_{2}}dp\tilde{u}(p)e^{pt}\nonumber\\
=&&\int_{p_{m}}^{p_{M}}\frac{-J/\sqrt{N}\sum_{k}\frac{1}{p+\omega_{k}+i0^{+}}e^{ipt}}{p+E_{1}-\sum_{k}\frac{J^{2}}{p+\omega_{k}+i0^{+}}}dp\nonumber\\
&&+\int_{p_{M}}^{p_{m}}\frac{-J/\sqrt{N}\sum_{k}\frac{1}{p+\omega_{k}-i0^{+}}e^{ipt}}{p+E_{1}-\sum_{k}\frac{J^{2}}{p+\omega_{k}-i0^{+}}}dp.\label{eq:B18}
\end{eqnarray}
Here,
\begin{eqnarray}
&&\sum_{k}\frac{1}{p+\omega_{k}\pm i0^{+}}\nonumber\\
=&&\frac{N}{2\pi}\int_{-\pi}^{\pi}dk\frac{1}{p+\omega_{k}\pm i0^{+}}\nonumber\\
=&&\frac{N}{2\pi}\left[\int_{-\pi}^{\pi}dkP\left(\frac{1}{p+\omega_{k}}\right)\mp\frac{i2\pi}{\sqrt{4\xi^{2}-(\omega_{0}+p)^{2}}}\right],\nonumber\\
\label{eq:B19}
\end{eqnarray}
where $P(x)$ is the principal value of $x$, and $\int_{-\pi}^{\pi}dkP[1/(p+\omega_{k})]=0$. Substituting Eq.~(\ref{eq:B19}) into Eq.~(\ref{eq:B18}) yields
\begin{eqnarray}
&&\int_{p_{m}}^{p_{M}}\frac{-\frac{J}{\sqrt{N}}\sum_{k}\frac{1}{p+\omega_{k}+i0^{+}}e^{ipt}}{p+E_{1}-\sum_{k}\frac{J^{2}}{p+\omega_{k}+i0^{+}}}dp\nonumber\\
&&+\int_{p_{M}}^{p_{m}}\frac{-\frac{J}{\sqrt{N}}\sum_{k}\frac{1}{p+\omega_{k}-i0^{+}}e^{ipt}}{p+E_{1}-\sum_{k}\frac{J^{2}}{p+\omega_{k}-i0^{+}}}dp\nonumber\\
=&&\frac{J}{\sqrt{N}}\int_{-2\xi}^{2\xi}\frac{\frac{2iN}{\sqrt{4\xi^{2}-(x)^{2}}}[(x-\omega_{0})+E_{1}]}{(E_{1}-\omega_{0}+x)^{2}+\frac{J^{2}N^{2}}{4\xi^{2}-(x)^{2}}}e^{i(x-\omega_{0})t}dx.\nonumber\\
\label{eq:B21}
\end{eqnarray}
Thus, we can obtain
\begin{eqnarray}
&&-\frac{1}{2\pi i}\left(\int_{l_{1}}dp\tilde{u}(p)e^{pt}+\int_{l_{2}}dp\tilde{u}(p)e^{pt}\right)\nonumber\\
&&=\int_{-2\xi}^{2\xi}C(x)e^{i(x-\omega_{0})t}dx,\label{eq:B22}
\end{eqnarray}
where
\begin{eqnarray}
C(x)=-\frac{1}{\pi}\frac{\frac{g}{\sqrt{4\xi^{2}-(x)^{2}}}[(x-\omega_{0})+E_{1}]}{(E_{1}-\omega_{0}+x)^{2}+\frac{J^{2}N^{2}}{4\xi^{2}-(x)^{2}}}e^{i(x-\omega_{0})t}.
\end{eqnarray}
In summary, we can write the expression of the probability amplitude of $|0,E_1\rangle$ as
\begin{eqnarray}
u(t)\!\!&=&\!\!B_{1}Q(p_{1})e^{p_{1}t}+B_{2}Q(p_{2})e^{p_{2}t}+\int_{-2\xi}^{2\xi}C(x)e^{i(x-\omega_{0})t}dx,\nonumber\\
\label{eq:B23}
\end{eqnarray}
where
\begin{eqnarray}
Q(p_{1})&=&\frac{g}{\xi\sqrt{M^{2}-4}},\label{eq:B23}\\
Q(p_{2})&=&-\frac{g}{\xi\sqrt{M^{2}-4}}, \label{eq:B24}\\
B_j&=&\frac{1}{\frac{df_{j}}{dp}|_{p_{j}}}\nonumber\\
   &=&\frac{(ip_{j}-\omega_{0})^{2}-4\xi^{2}}{(ip_{j}-\omega_{0})^{2}-4\xi^{2}+(ip_{j}-E_{1})(ip_{j}-\omega_{0})}.\nonumber\\
   \label{eq:B25}
\end{eqnarray}

\bibliography{ref}

\end{document}